\documentclass[3p, review, times, compress]{elsarticle}  
\usepackage{amssymb}
\usepackage{amsmath}
\usepackage[colorlinks=true,linkcolor=black,citecolor=blue,urlcolor=blue,]{hyperref}
\usepackage{multirow} 
\usepackage{caption}
\setcitestyle{open={[},close={]},citesep={,\!},numbers}

\journal{Journal of Energy Storage}

\begin{document}

\renewcommand{\figureautorefname}{Fig.}
\renewcommand{\sectionautorefname}{Section}
\captionsetup[figure]{labelfont={bf},name={Fig.},labelsep=period}

\begin{frontmatter}

\title{A multi-scale lithium-ion battery capacity prediction using mixture of experts and patch-based MLP}

\author{Yuzhu Lei}
\author{Guanding Yu\corref{mycorrespondingauthor}}
\cortext[mycorrespondingauthor]{Corresponding author}
\ead{yuguanding@zju.edu.cn}

\address{College of Information Science and Electronic Engineering, Zhejiang University, Hangzhou 310027, China}

\begin{abstract}
Lithium-ion battery health management has become increasingly important as the application of batteries expands. Precise forecasting of capacity degradation is critical for ensuring the healthy usage of batteries. In this paper, we innovatively propose MSPMLP, a multi-scale capacity prediction model utilizing the mixture of experts (MoE) architecture and patch-based multi-layer perceptron (MLP) blocks, to capture both the long-term degradation trend and local capacity regeneration phenomena. Specifically, we utilize patch-based MLP blocks with varying patch sizes to extract multi-scale features from the capacity sequence. Leveraging the MoE architecture, the model adaptively integrates the extracted features, thereby enhancing its capacity and expressiveness. Finally, the future battery capacity is predicted based on the integrated features, achieving high prediction accuracy and generalization. Experimental results on the public NASA dataset indicate that MSPMLP achieves a mean absolute error (MAE) of 0.0078, improving by 41.8\% compared to existing methods. These findings highlight that MSPMLP, owing to its multi-scale modeling capability and generalizability, provides a promising solution to the battery capacity prediction challenges caused by capacity regeneration phenomena and complex usage conditions. The code of this work is provided at \href{https://github.com/LeiYuzhu/CapacityPredict}{https://github.com/LeiYuzhu/CapacityPredict}.
\end{abstract}


\end{frontmatter}

\section{Introduction}
Increased demand for electronic products, electric tools, transportation, aerospace, and medical devices has led to a surge in the usage of lithium-ion batteries \cite{r1,r2,r16}. However, lithium-ion batteries experience performance decline and capacity reduction during repeated cycling \cite{r10,r48}. These issues primarily stem from complex internal electrochemical reactions and physical changes within the batteries, such as solid electrolyte interphase (SEI) film growth and active material loss \cite{r37,r38}. Predicting battery capacity variations aids in evaluating battery state and optimizing charging-discharging strategies. Furthermore, accurate capacity prediction can mitigate battery safety incidents, enhancing battery safety and reliability \cite{r4,r36,r53}. Concurrently, precise capacity forecasting supports battery recycling and reuse, fostering sustainable resource utilization. Therefore, battery capacity prediction holds paramount significance in the 
full life cycle management of lithium-ion batteries \cite{r9,r47}. 

However, the actual capacity degradation of lithium-ion batteries is intricate, as the existence of regeneration phenomena can lead to abrupt changes in the capacity curves \cite{r21}. Specifically, the lithium-ion battery capacity degradation exhibits an overall decreasing trend with multiple fluctuations. Consequently, predicting future capacity changes remains a challenging task \cite{r50,r3,r11}. To mitigate the effects of capacity regeneration, He et al. decouple the battery capacity sequences into global degradation and local fluctuations by wavelet analysis \cite{r3}. Subsequently, Gaussian process regression (GPR) is employed to model the extracted degradation and fluctuations, yielding superior results compared to the GPR model without decoupling. Similarly, in \cite{r11}, the empirical mode decomposition (EMD) is applied to decompose the capacity sequence into a low-frequency residual and multiple intrinsic mode functions (IMFs), representing the global degradation trend and local fluctuations, respectively. Multiscale logic regression and GPR are then utilized to estimate these components. Furthermore, various models, including long short-term memory (LSTM)\cite{r14,r41,r30,r44}, temporal convolutional network (TCN) \cite{r13}, Elman neural network \cite{r46,r55} and gated recurrent unit (GRU) \cite{r19}, have also been integrated to fit the residual and IMFs. These decomposition-based methods mitigate the impact of local capacity regeneration phenomena by isolating fluctuations. However, decomposition and separate estimation inevitably lead to information loss. Besides, estimation errors from individual components accumulate during the reconstruction process, reducing the overall accuracy. It is imperative to develop models that can more effectively integrate local feature extraction and global dependency modeling.

In addition, variations in battery manufacturing, environmental and operating conditions affect the decline of battery capacity \cite{r22,r17,r49}. Consequently, the capacity degradation process is highly cell-specific \cite{r25,r15,r39}. For practical utility, the generalization is also a key consideration of the battery capacity prediction model \cite{r5,r7,r8}. Some studies propose to improve the capacity prediction performance by combining multiple base prediction models. For example, Wu et al. integrate five distinct models to improve the accuracy and generalization of the time series prediction model \cite{r20}. However, ensemble models often introduce significant complexity, posing challenges to model debugging and interpretability. In \cite{r18}, utilizing mixture of experts (MoE), the prediction results of multiple networks are combined to enhance the robustness of the prediction model. Nonetheless, the efficacy of simple prediction result fusion remains limited. To achieve superior generalization, informative and adaptive feature extraction is required to enhance the overall modeling capability.

In recent years, models based on multi-layer perceptron (MLP) have made significant progress in computer vision, including MLP-Mixer \cite{r35} and ResMLP \cite{r34}. A key innovation in these MLP-based vision models lies in the introduction of the patch mechanism, drawing inspiration from the image partitioning strategy of vision transformer (ViT) and the regional feature learning concept of convolutional neural network (CNN). The patch mechanism allows for flexible adjustment of image patch sizes, enabling the model to effectively extract local image details. Results show that these MLP-based models can achieve performance comparable to ViT while simplifying the complexity. Inspired by this, the field of time series prediction has also begun to pay attention to MLP-based models. Tang et al. propose a prediction model incorporating patch embedding and feature decomposition. In this approach, the embedded features are decomposed into smooth components and noise residuals, allowing for separate feature extraction by MLPs \cite{r23}. Additionally, Zhang et al. utilize multiple MSD-Mixer layers with different patch sizes to model multi-scale temporal patterns in sequences \cite{r24}. In the MSD-Mixer layer, the sequence is first divided into multiple patches, and then intra-patch and inter-patch features are captured by MLPs. Research findings suggest that these MLP-based models exhibit superior performance compared to widely adopted Transformer models when processing time series with noise and redundant features.

Some researchers have also attempted to apply MLP-based architectures to forecast the battery capacity. Zhao et al. cascade BiGRU, attention mechanism module, MLP-Mixer and MoE to capture sequence features, so as to provide more reliable and accurate prediction results of battery remaining life \cite{r32}. Similarly, Wu et al. capture time series features and predict the battery remaining life by cascading frequency domain attention transformer, BiLSTM, multi-head attention module, DualMLP and MoE predictor \cite{r33}. However, these simple cascading of multiple feature extraction models greatly increase the complexity and the inference time, which are inefficient and lack interpretability as well. Ye et al. propose an MLP-Mixer-based architecture to recognize long-term trends and local details within time series, but the single fixed patch size adopted restricts the generalization and leads to limited prediction performance \cite{r51}. Designing an effective MLP-based model for battery capacity degradation prediction is still a research direction worthy of attention.

Inspired by the above studies, this paper designs an innovative model that integrates the MoE architecture and patch-based MLP blocks. The significant contributions of this research are detailed below:

(1) We introduce a novel patch-based MLP block design that facilitates the extraction of both local fluctuations and long-term trends in battery capacity degradation by segmenting time series into patches and utilizing MLPs for both intra-patch and inter-patch feature extraction. This architecture enables the model to more precisely extract time-series features and predict battery health. Furthermore, the flexibility to adjust patch sizes allows the model to adapt to varying temporal scales of degradation and regeneration, ensuring robust performance across diverse battery data.

(2) To capture multi-scale characteristics and enhance the generalization, we employ MoE to combine features from patch-based MLP blocks with varying patch sizes. Through the MoE architecture, adaptive selection of time scales (patch sizes) is achieved using the gating network, overcoming the limitations of fixed time-scale feature extraction in traditional methods. Furthermore, we flexibly activate the top-k experts with the highest weights to significantly reduce redundancy and enhance computational efficiency by focusing on the most relevant experts. By adaptively prioritizing experts based on their contribution, we ensure that the model remains lean and accurate, effectively addressing the challenges of multi-scale feature integration in complex datasets.

(3) Trained and tested on batteries under different operating conditions, MAPMLP shows robust generalization and high accuracy. Results demonstrate that the mean absolute error (MAE) of battery capacity forecasting achieves 0.0078, which improves by 41.8\% compared to existing advanced methods. This significant performance enhancement fully demonstrates the generalization and accuracy of MSPMLP in handling complex battery degradation data. It can effectively address the challenges posed by capacity regeneration phenomena and complex operating conditions, providing a reliable solution for battery health management.

The structure of the remaining sections is presented below. \autoref{Section 2} introduces the problem formulation and model framework. Then, \autoref{Section 3} describes the experimental settings, including the dataset, baselines, evaluation metrics and parameter settings. Subsequently, \autoref{Section 4} presents a comprehensive analysis of the experimental results, exhibiting the performance of MSPMLP in comparison to baselines and the impact of each module of MSPMLP. Finally, primary findings and prospective research avenues are concluded in \autoref{Section 5}.

\section{Methodology} \label{Section 2}
\begin{figure*}[t]
	\vspace{-2ex}
	\centering
	\begin{minipage} {0.32 \linewidth}
		\centering
	    \includegraphics[width=1\textwidth]{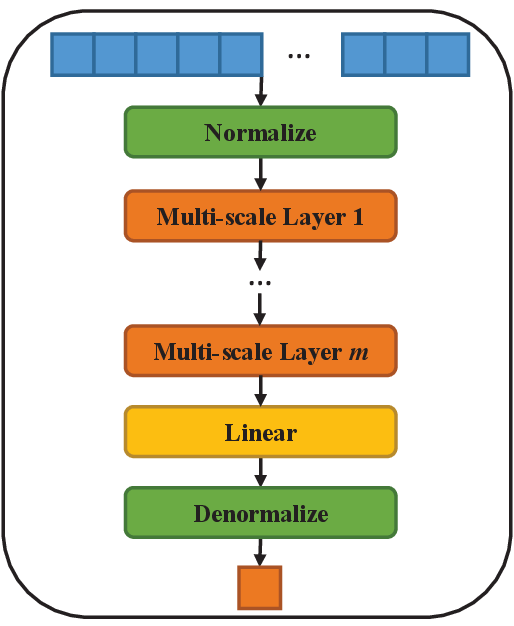}
	    \caption{MSPMLP overview.}
	    \label{fig_1}
	\end{minipage}
	\begin{minipage} {0.63 \linewidth}
		\centering
	    \includegraphics[width=1\textwidth]{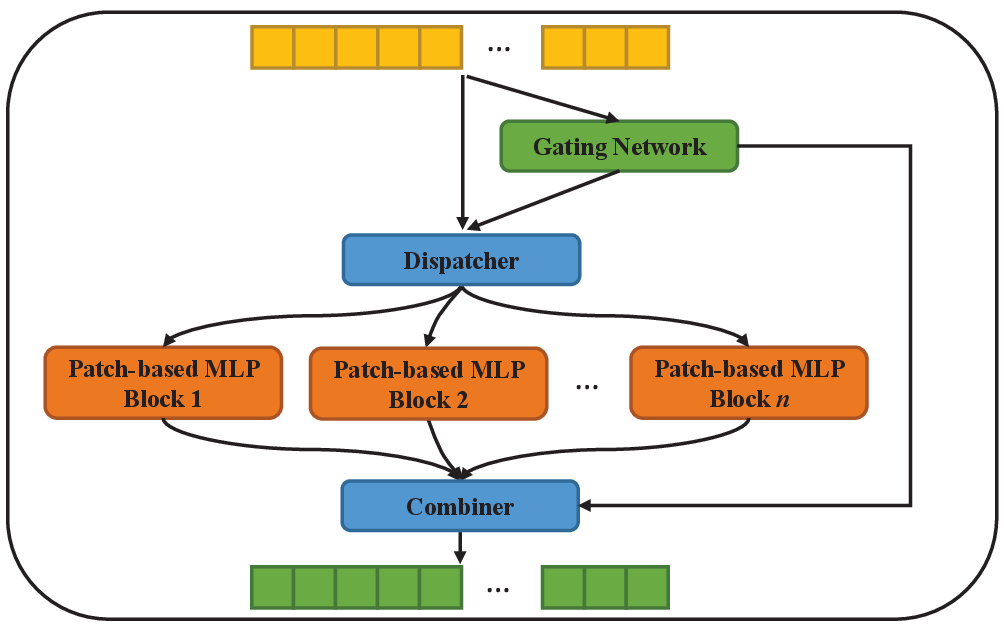}
	    \caption{Multi-scale layer overview.}
	    \label{fig_2}
	\end{minipage}
	\vspace{-2ex}
\end{figure*}

\subsection{Problem Formulation}
Define the capacity of battery $i$ at its $j$ th cycle as $c_{i,j}$. The capacity prediction problem can be formally described as 
\begin{equation}
\hat{c}_{i,t+1}=f([c_{i,t-w+1}, c_{i,t-w+2},... , c_{i,t}])
\end{equation}
where $\hat{c}_{i,t+1}$ represents the predicted capacity of battery $i$ at the $t+1$ th cycle, $[c_{i,t-w+1}, c_{i,t-w+2},... , c_{i,t}]$ denotes the historically observed capacity sequence, $w$ represents the observation window length, and $f$ is the mapping function to be established.

\subsection{MSPMLP overview}
\autoref{fig_1} illustrates the framework of MSPMLP. Specifically, the MSPMLP model consists of a normalization layer, $m$ multi-scale layers, a linear prediction layer and a denormalization layer. The historically observed capacity sequence $[c_{i,t-w+1}, c_{i,t-w+2},... , c_{i,t}]$ is first normalized to improve the performance and stability. Subsequently, multi-scale layers are utilized to capture the multi-scale features. The cascade of $m$ multi-scale layers can provide more refined feature representations and better adapt to various changes in the sequence, thus improving the prediction accuracy and model robustness. Finally, the extracted features are fed into the linear layer and the denormalization layer to obtain the predicted capacity.

\subsection{Mixture of experts architecture}
As illustrated in \autoref{fig_2}, the multi-scale layer is designed based on the MoE architecture. Specifically, each multi-scale layer consists of a gating network, a dispatcher, a combiner and $n$ independent experts, i.e., the patch-based MLP blocks. The input is fed into the gating network and all experts. The gating network outputs a probability distribution representing the weights of experts. Then the experts with different patch sizes extract features of different time scales in parallel. Finally, the extracted features are combined according to the weights to obtain multi-scale features. In contrast to the multi-layer feature extraction with fixed time scales in \cite{r24,r51}, the gating network can provide an adaptive selection of time scales. For diverse input capacity sequences, this approach improves the flexibility and practicality. Besides, by combining multiple patch-based MLP blocks that focus on different scales, the expressive ability of the multi-scale layer is significantly enhanced. In addition, the intensive investigation of expert selection can also reveal the decision-making pathways of the model, contributing to enhanced interpretability. 

However, excessively increasing the number of experts to enhance the diversity of scales may introduce redundant information and reduce prediction accuracy. In this paper, we choose to flexibly activate the top-k experts with the highest weights. Through the adaptive selection of scales informed by the input sequence characteristics, this mechanism enables precise feature capture and improves the accuracy. Meanwhile, selectively activating experts also reduces redundant computations and lowers the demand for computational resources.

\begin{figure}[t]
\centering
\includegraphics[width=5.5in]{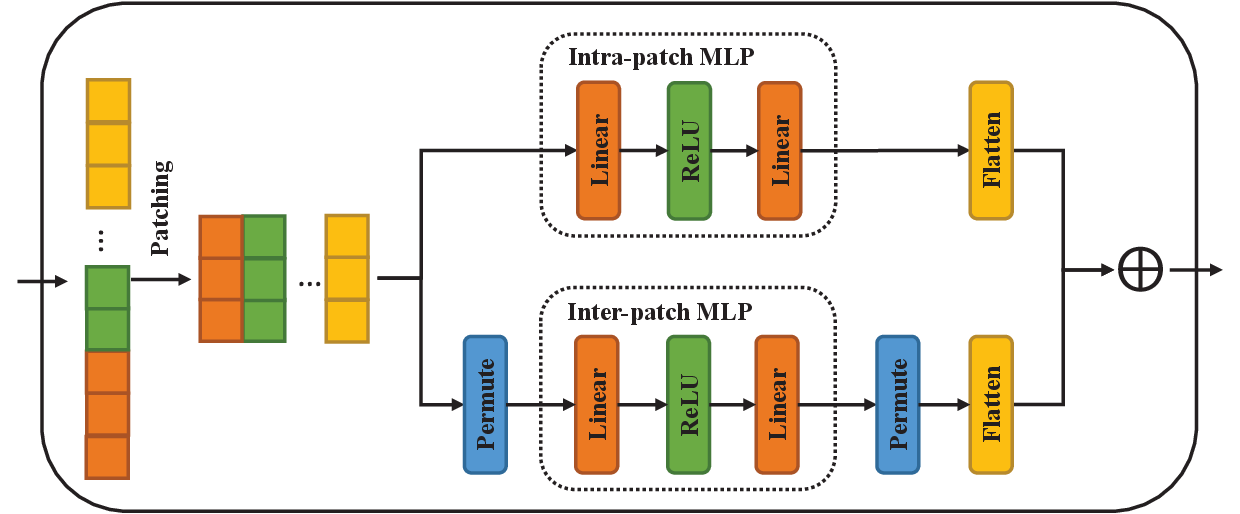}
\caption{Patch-based MLP block overview.}
\label{fig_3}
\end{figure}

A key challenge inherent in the selective activation strategy arises from the heterogeneity of features within a parallel computation batch, which results in varying expert selections across constituent sequences. We employ a dispatcher to overcome this problem. Based on the top-k experts selected by each sequence within the batch, we efficiently partition the input batch into multiple sub-batches. Subsequently, each expert processes its corresponding sub-batch in parallel. Finally, the outputs from all experts are concatenated and combined through element-wise multiplication with the weights. This approach enhances the robustness for parallel processing of large and diverse sequence data and accelerates model training and inference.

\subsection{Patch-based MLP Block}
The architecture of the patch-based MLP block is shown in \autoref{fig_3}. Assuming the input of the patch-based MLP block is $X=[x_1, x_2, ..., x_w]$. After the patching operation with a patch size of $p$, the input $X$ is divided into $N$ patches, denoted as $X_{p}=[P_1, P_2, ..., P_N]$, where $N=w/p$. Note that $p$ is a parameter that can be adjusted according to the length and characteristics of the capacity sequence. Subsequently, we use intra-patch MLP and inter-patch MLP to capture the relationships between the time steps within each patch and the relationships between patches, respectively. Specifically, $X_{p}$ is input into an MLP with $l_{intra}$ linear layers and $d_{intra}$ hidden neurons per layer to obtain the local features in all patches. Then, through the flatten operation, we get the local details of the input sequence. As for the inter-patch features, $X_{p}$ is first permuted and then input into an MLP with $l_{inter}$ linear layers and $d_{inter}$ hidden neurons per layer to obtain the dependencies between patches. By permuting again and flattening, we get the global dependencies in the input sequence. We finally fuse the intra-patch features and inter-patch features to obtain the features with the patch size $p$. By varying the patch size, the model can effectively capture local details and long-term dependencies across different temporal scales, thereby extracting multi-scale features and enhancing the expressive capacity.

\begin{figure}[t]
	\vspace{-2ex}
	\centering
	\begin{minipage} {0.49 \linewidth}
		\centering
	    \includegraphics[width=1\textwidth]{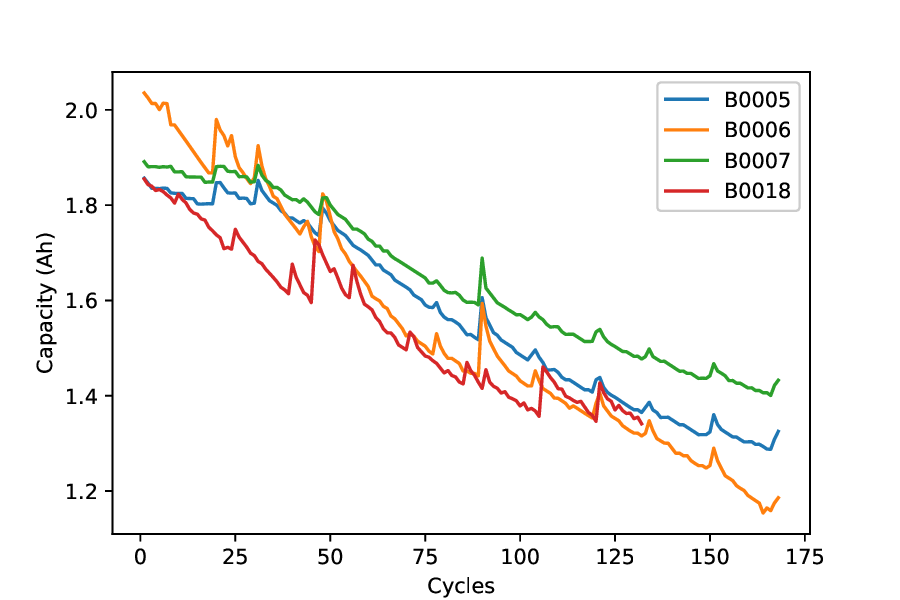}
	    \vspace{-5ex}
	    \caption{The capacity degradation of NASA dataset.}
	    \label{fig_4}
	\end{minipage}
	\begin{minipage} {0.49 \linewidth}
		\centering
	    \includegraphics[width=1\textwidth]{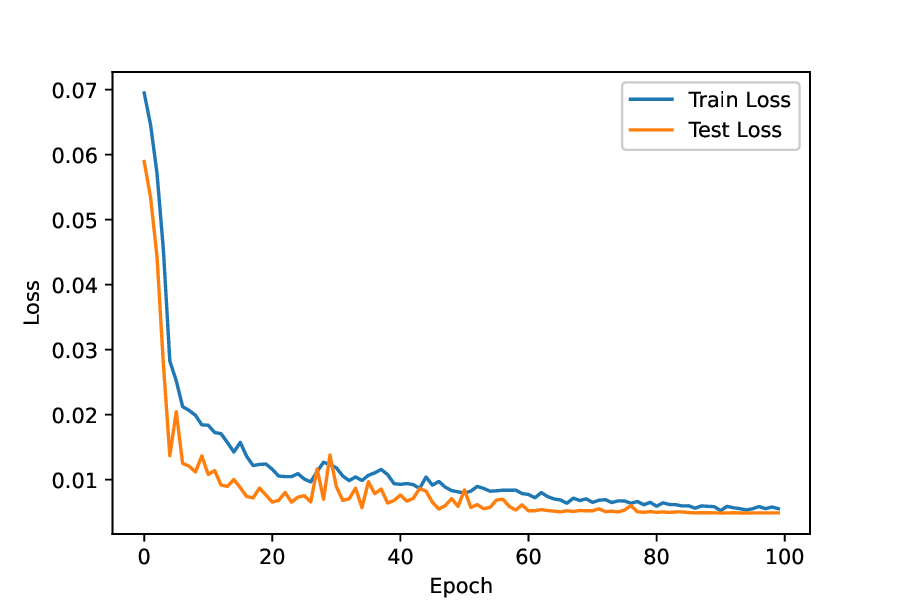}
	    \vspace{-5ex}
	    \caption{Train loss and test loss of B0005.}
	    \label{fig_5}
	\end{minipage}
	\vspace{-2ex}
\end{figure}

\section{Experimental settings} \label{Section 3}
\subsection{Dataset}
Utilizing the public NASA dataset \cite{r26}, we evaluate the MSPMLP model on aging data from four batteries (B0005, B0006, B0007, and B0018) cycled under three distinct operational settings. \autoref{fig_4} demonstrates the changes in battery capacity as charge-discharge cycles increase. Obviously, the capacity degradation of the four batteries is accompanied by various capacity regenerations, posing challenges for capacity prediction.

The sliding window approach is utilized to convert the battery capacity data into supervised learning datasets \cite{r27}. To validate the generalization and practical applicability of MSPMLP, this work employs the leave-one-out cross-validation (LOOCV) strategy \cite{r54}. In contrast to studies employing early-stage battery data for training \cite{r14,r41,r30}, the data from the remaining three batteries is used for training, enabling early capacity prediction.

\subsection{Baselines}
We conduct a comprehensive comparative analysis of our method by evaluating its performance against several advanced baselines, including deep neural network (DNN), MoE, Transformer\cite{r31,r52}, ASTLSTM\cite{r28,r29}, AttMoE\cite{r18}, and EMD+LSTM+GPR\cite{r30}.

\begin{itemize}
\item DNN: Compared to MLP, DNN learns more complex feature representations by increasing the depth of the network, i.e., adding more hidden layers.

\item MoE: In this method, simple MLP models, instead of the proposed patch-based MLP blocks, are employed as experts. By combining multiple experts, it is expected that the capacity and expressiveness of the model can be enhanced.

\item Transformer: In Transformer, the multiple attention heads facilitate the parallel extraction of sequence features from multiple perspectives. It can obtain comprehensive feature representations and is a widely adopted method for time series prediction.

\item ASTLSTM: Different from standard LSTMs, ASTLSTM utilizes a fixed connection between input and forget gates to jointly select the old information and new data. It further refines information retention through an element-wise product of input and past cell state.

\item AttMoE: By cascading multi-head attention module and MoE predictor with fully connected layers as experts, AttMoE aims to effectively capture degradation features and enhance the generalization capability.

\item EMD+LSTM+GPR: EMD is used to decompose the original battery capacity data. The LSTM is employed to estimate the global trend, while the GPR is utilized to fit various fluctuations. Therefore, both the long-term dependency and the local capacity regeneration can be captured.
\end{itemize}

\subsection{Evaluation metrics}
We employ MAE and root mean squared error (RMSE) to evaluate the efficacy of capacity prediction methods, i.e.,

\begin{equation}
MAE=\frac{1}{T}\sum\limits_{t=1}^T | c_{i,t}-\hat{c}_{i,t} |
\end{equation}
\begin{equation}
RMSE=\sqrt{\frac{1}{T}\sum\limits_{t=1}^T ( c_{i,t}-\hat{c}_{i,t} )^2}
\end{equation}
where $T$ denotes the number of capacity predictions.

\begin{table} 
\centering
\begin{tabular}{l c c c c c c c r}
\hline
Battery & Metric & DNN & MoE & Transformer & ASTLSTM & AttMoE & EMD+LSTM+GPR & Proposed \\
\hline
\multirow{2}*{B0005} & MAE & 0.0150 & 0.0148 & 0.0097 & 0.0082 & 0.0097 & 0.0087 & \pmb{0.0046} \\
& RMSE & 0.0207 & 0.0206 & 0.0152 & 0.0147 & 0.0166 & 0.0115 & \pmb{0.0105} \\
\multirow{2}*{B0006} & MAE & 0.0313 & 0.0314 & 0.0225 & 0.0180 & 0.0188 & 0.0260 & \pmb{0.0086} \\
& RMSE & 0.0382 & 0.0384 & 0.0294 & 0.0261 & 0.0278 & 0.0341 & \pmb{0.0187} \\
\multirow{2}*{B0007} & MAE & 0.0246 & 0.0240 & 0.0205 & 0.0121 & 0.0118 & 0.0125 & \pmb{0.0044} \\
& RMSE & 0.0306 & 0.0297 & 0.0254 & 0.0175 & 0.0180 & 0.0153 & \pmb{0.0109} \\
\multirow{2}*{B0018} & MAE & 0.0245 & 0.0264 & 0.0229 & 0.0155 & 0.0156 & 0.0144 & \pmb{0.0136} \\
& RMSE & 0.0330 & 0.0353 & 0.0331 & 0.0281 & 0.0258 & \pmb{0.0214} & 0.0260 \\
\hline
\end{tabular}
\caption{MAE and RMSE of capacity prediction for different methods on the four batteries.}\label{tab_1}
\end{table}

\begin{figure}[t]
\centering
\vspace{-2ex}
\includegraphics[width=6.6in]{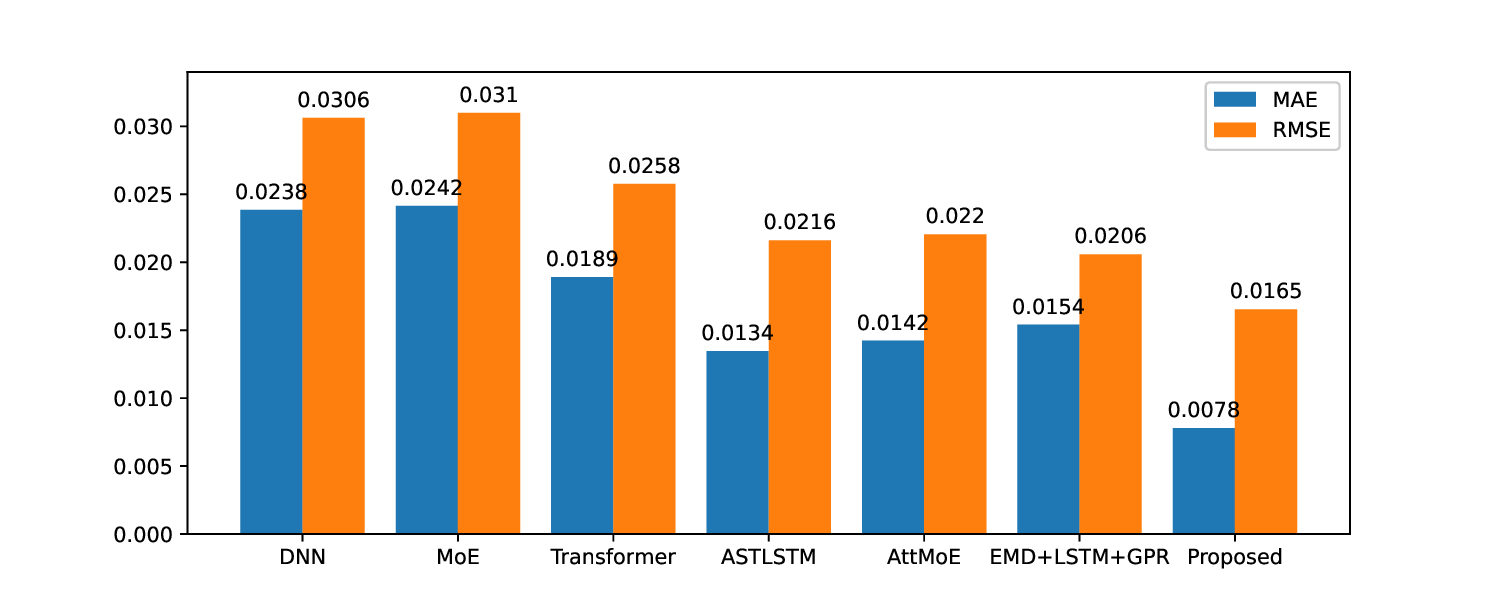}
\vspace{-6ex}
\caption{Average MAE and RMSE of capacity prediction for different methods on the four batteries.}
\label{fig_6} 
\end{figure}

\begin{figure}[t]
	\vspace{-2ex}
	\centering
	\begin{minipage} {0.49 \linewidth}
		\centering
	    \includegraphics[width=1\textwidth]{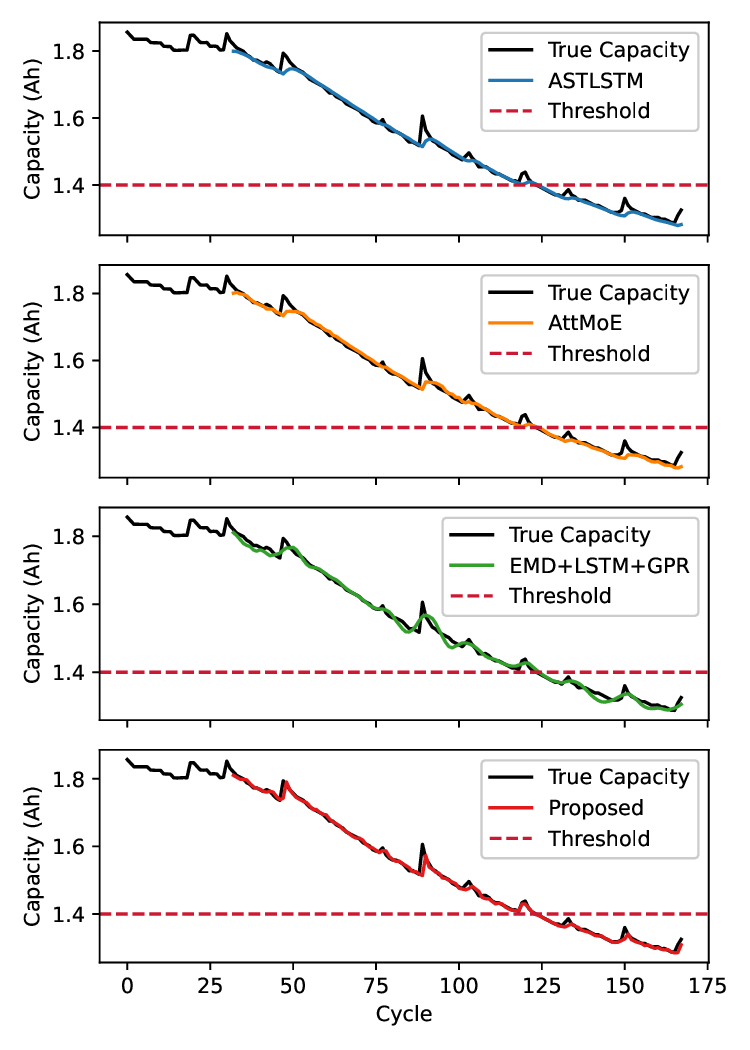}
	    \vspace{-6ex}
	    \caption{Capacity prediction results of B0005.}
	    \label{fig_7}
	\end{minipage}
	\begin{minipage} {0.49 \linewidth}
		\centering
	    \includegraphics[width=1\textwidth]{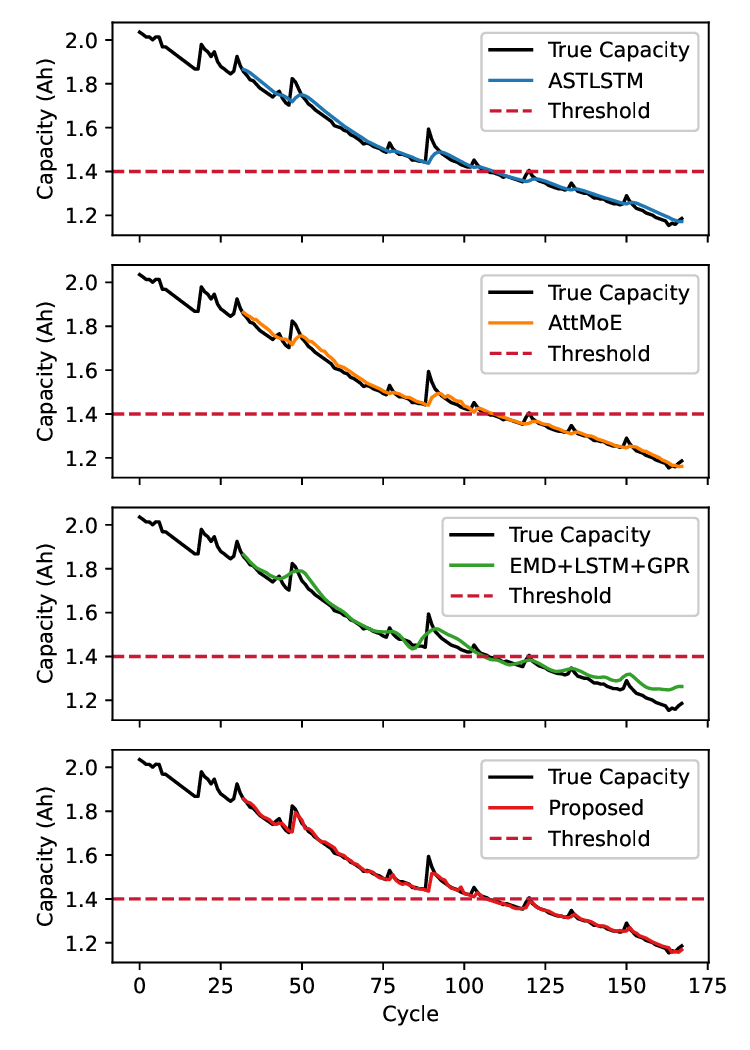}
	    \vspace{-6ex}
	    \caption{Capacity prediction results of B0006.}
	    \label{fig_8}
	\end{minipage}
	\vspace{-2ex}
\end{figure}

\begin{figure}[t]
	\vspace{-2ex}
	\centering
	\begin{minipage} {0.49 \linewidth}
		\centering
	    \includegraphics[width=1\textwidth]{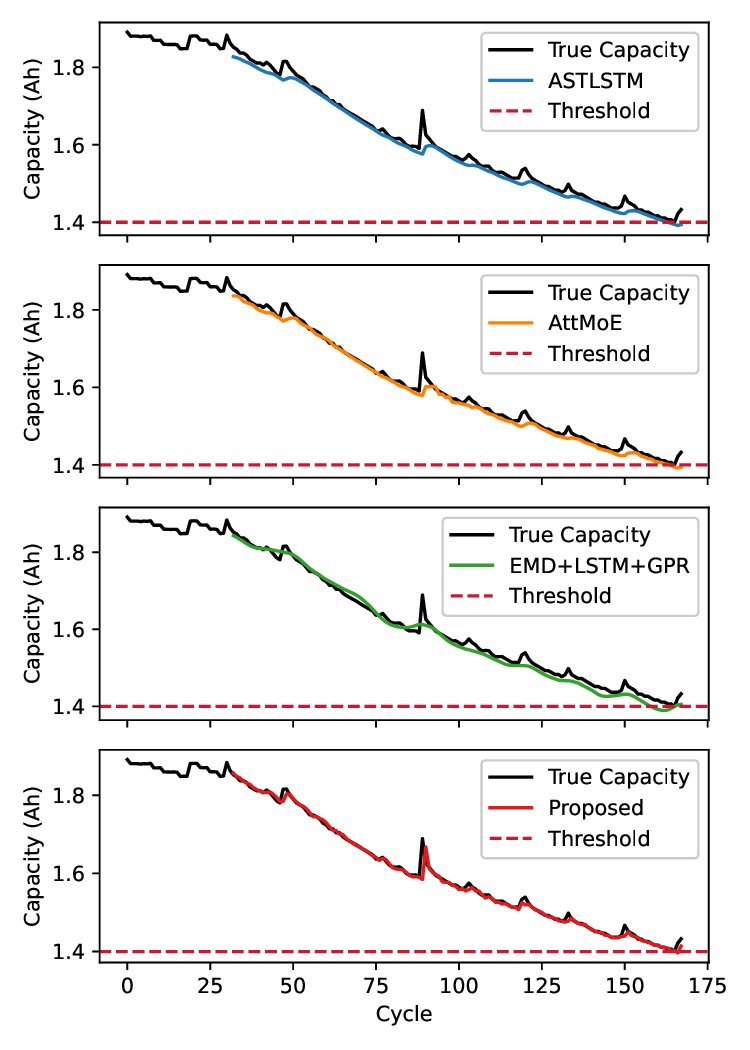}
	    \vspace{-6ex}
	    \caption{Capacity prediction results of B0007.}
	    \label{fig_9}
	\end{minipage}
	\begin{minipage} {0.49 \linewidth}
		\centering
	    \includegraphics[width=1\textwidth]{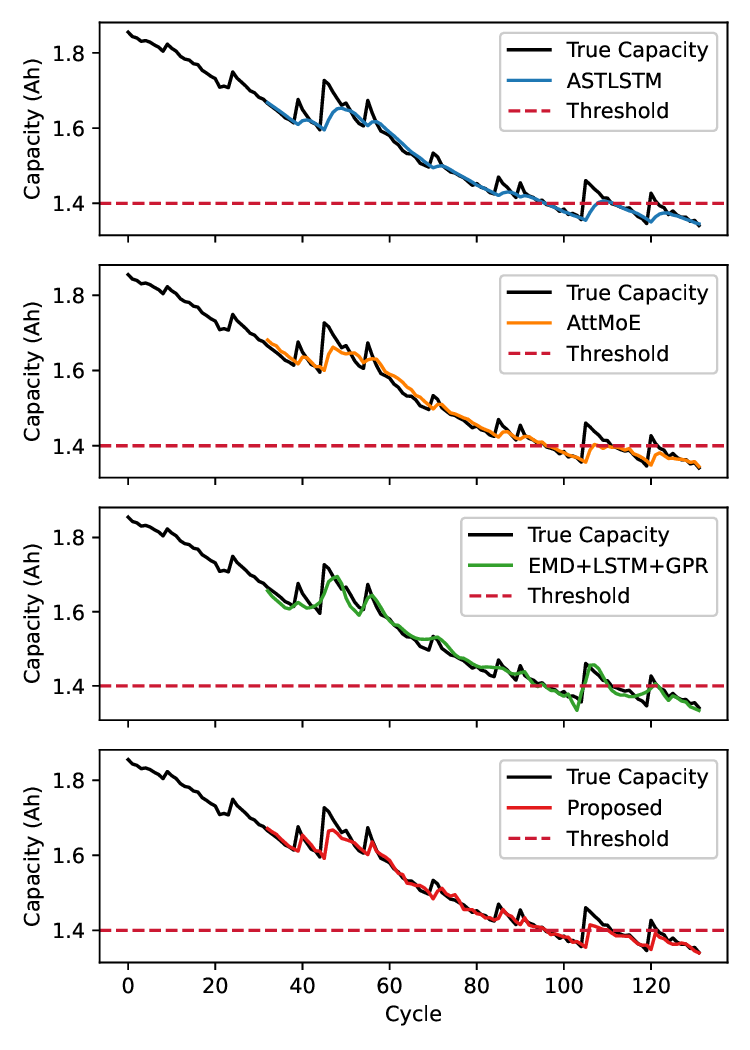}
	    \vspace{-6ex}
	    \caption{Capacity prediction results of B0018.}
	    \label{fig_10}
	\end{minipage}
	\vspace{-2ex}
\end{figure}

The two metrics assess the performance of capacity prediction models by comparing the predicted capacity with the actual capacity. Generally, MAE provides the average magnitude of errors, while RMSE is more sensitive to larger errors. Therefore, the prediction performance can be comprehensively evaluated by analyzing and comparing the MAE and RMSE metrics, allowing for the assessment of both overall error magnitude and the prediction accuracy within capacity regeneration regions, which are prone to significant prediction errors.

\subsection{Parameter settings}
Based on experience and experiments, the length of historical observation window is set to $w = 36$, the number of multi-scale layers is determined to be $m = 2$, and the number of experts in each multi-scale layer is determined to be $n = 4$. The patch sizes for the experts in the two multi-scale layers are configured as $[18, 12, 9, 6]$ and $[6, 4, 3, 2]$, respectively. For B0005, B0006 and B0007, the number of experts to be selectively activated is set to $k = 3$. However, $k$ is set to 1 for B0018. Additionally, the number of linear layers in both the intra-patch MLP and inter-patch MLP is set to $l_{intra}=l_{inter}=2$, and the number of hidden neurons is set to $d_{intra}=d_{inter}=64$. Subsequently, the model is trained for 100 epochs with a batch size of 32, utilizing MAE as the loss function and Adam as the optimizer with a learning rate of 0.005. To guarantee result reliability, the model is evaluated based on the average of battery performance metrics obtained from five independent trials.

\section{Results and discussion} \label{Section 4}
\subsection{Convergence}
\autoref{fig_5} illustrates the training and testing loss curves using B0005 as the test set and B0006, B0007, and B0018 as the training set. The loss function exhibits a rapid decline within the initial 20 training epochs, subsequently stabilizing with minor oscillations. Both training and testing losses demonstrate consistent trends, ultimately converging below 0.01, suggesting that the model has achieved convergence without overfitting or underfitting. These results indicate that the model has effectively reconciled the trade-off between training data fidelity and inherent generalization capacity during the training phase, facilitating robust knowledge transfer from the training set to the test set.

\subsection{Capacity prediction performance}
Capacity prediction performance for the four batteries is tabulated in \autoref{tab_1}, and the average MAE and RMSE are graphically illustrated in \autoref{fig_6}. The results demonstrate that MSPMLP achieves the best battery capacity prediction performance, with an average MAE of 0.0078, representing a 41.8\% improvement over the second-best method, ASTLSTM. This demonstrates the effectiveness of MAPMLP leveraging the MoE architecture and patch-based MLP blocks to capture multi-scale features. Moreover, the significant performance improvement compared to DNN and MoE methods demonstrates the effectiveness of the patch mechanism in capturing multi-scale features and enhancing model expressiveness. This confirms the validity of the model structure specifically designed for the characteristics of battery capacity degradation. Additionally, our method achieved an average RMSE of 0.0165, lower than other methods, but less improved than MAE. This indicates that MSPMLP still exhibits some relatively large errors in capacity regeneration regions, especially on battery B0018, which experienced multiple larger capacity regenerations.

To visually clarify the capacity prediction performance, \autoref{fig_7}, \autoref{fig_8}, \autoref{fig_9}, and \autoref{fig_10} present the capacity forecasting results of ASTLSTM, AttMoE, EMD+LSTM+GPR, and our proposed method on the four batteries, respectively. The results reveal that ASTLSTM and AttMoE have limited capabilities in capturing capacity regeneration phenomena. The EMD+LSTM+GPR method, benefiting from mode decomposition, can capture the global dependencies and capacity regeneration phenomena, showing better performance on most test points. However, the information loss inherent in mode decomposition and the limited generalization ability lead to deviations between the predicted and actual capacities. In contrast, our proposed method presents highly accurate predictions for batteries B0005, B0006, and B0007. For battery B0018, due to the lack of similar capacity regeneration trends in the training data (B0005, B0006, and B0007), the prediction accuracy is slightly lower. However, it still performs significantly better than ASTLSTM and AttMoE and obtains the best MAE. Moreover, by utilizing data from other batteries instead of the early-stage data of the target battery for training, early capacity prediction is feasible. Experimental results demonstrate that MSPMLP also offers high accuracy and strong robustness in early prediction.


\begin{figure}[t]
	\vspace{-2ex}
	\centering
	\begin{minipage} {0.49 \linewidth}
		\centering
	    \includegraphics[width=1\textwidth]{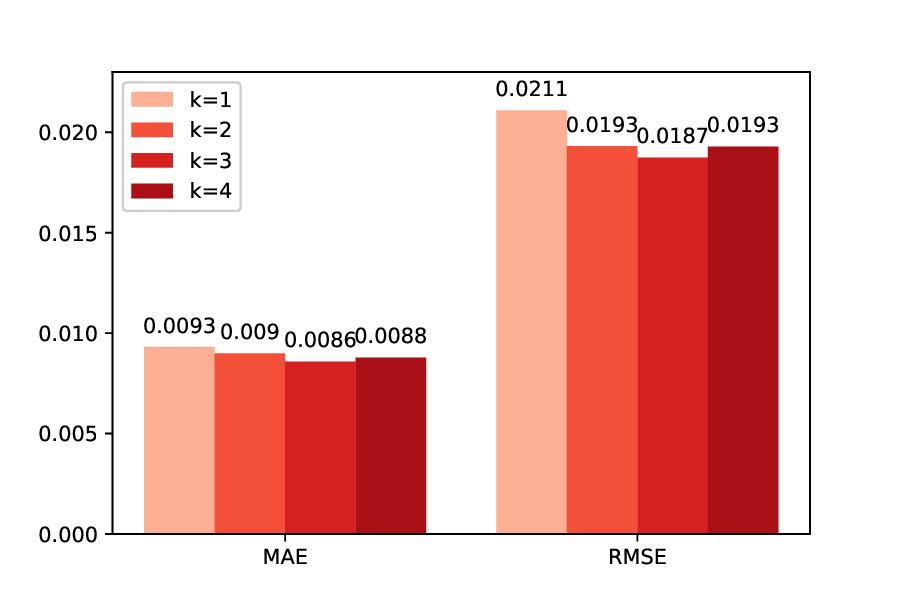}
	    \vspace{-6ex}
	    \caption{MAE and RMSE of capacity prediction with different numbers of activated experts on B0006.}
	    \label{fig_11}
	\end{minipage}
	\begin{minipage} {0.49 \linewidth}
		\centering
	    \includegraphics[width=1\textwidth]{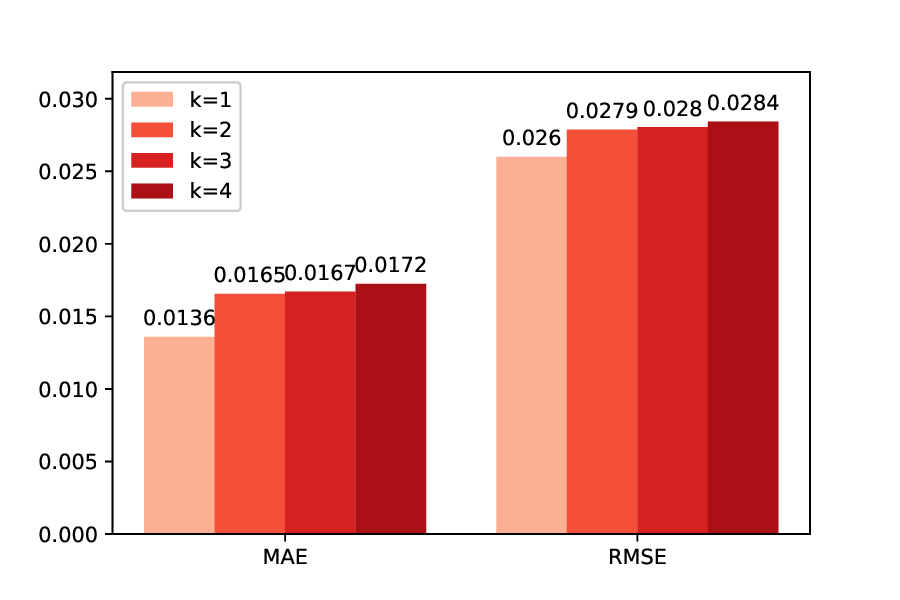}
	    \vspace{-6ex}
	    \caption{MAE and RMSE of capacity prediction with different numbers of activated experts on B0018.}
	    \label{fig_12}
	\end{minipage}
	\vspace{-2ex}
\end{figure}

\subsection{Effects of selective activation}
To evaluate the impact of the proposed selective activation strategy, we test the capacity prediction performance under different numbers of activated experts. \autoref{fig_11} and \autoref{fig_12} show the MAE and RMSE of B0006 and B0018 with different numbers of activated experts, respectively. For B0005, B0006 and B0007, the capacity regenerations are less frequent and relatively smaller in magnitude. The three batteries have similar performance with varying numbers of activated experts, thus we only present the result of B0006.

\begin{table} 
\centering
\begin{tabular}{l c c c r}
\hline
Battery & Metric  & Without intra-patch MLP & Without inter-patch MLP & Proposed \\
\hline
\multirow{2}*{B0005} & MAE & 0.0056 & 0.0048 & 0.0046 \\
  & RMSE & 0.0119 & 0.0108 & 0.0105 \\
\multirow{2}*{B0006} & MAE & 0.0099 & 0.0088 & 0.0086 \\
  & RMSE & 0.0212 & 0.0196 & 0.0187 \\
\multirow{2}*{B0007} & MAE & 0.0052 & 0.0045 & 0.0044 \\
  & RMSE & 0.0117 & 0.0110 & 0.0109 \\
\multirow{2}*{B0018} & MAE & 0.0150 & 0.0140 & 0.0136 \\
  & RMSE & 0.0272 & 0.0262 & 0.0260 \\
\hline
\end{tabular}
\caption{MAE and RMSE of capacity prediction on models with intra-patch MLP and inter-patch MLP or not.}\label{tab_2}
\end{table}

\begin{figure}[t]
\centering
\vspace{-2ex}
\includegraphics[width=3.4in]{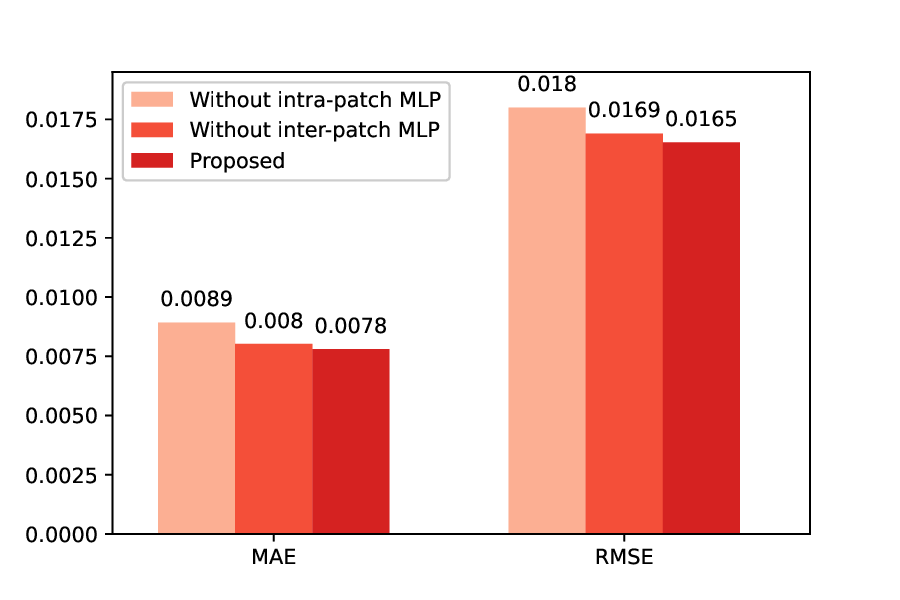}
\vspace{-3ex}
\caption{Average MAE and RMSE of capacity prediction on models with intra-patch MLP and inter-patch MLP or not.}
\label{fig_13} 
\end{figure}

The results indicate that the MAE of capacity prediction for B0006 improves by 8.1\% when increasing the number of activated experts from $k=1$ to $k=3$. Similarly, the RMSE of capacity prediction exhibits a 12.8\% improvement from $k=1$ to $k=3$. The improvement in RMSE is greater than that of MAE when increasing $k$ from 1 to 3. This indicates that with the increase in the number of activated experts, larger prediction errors can be effectively avoided. Therefore, in this case, increasing the number of selectively activated experts enhances the ability to perceive details and capture capacity regeneration phenomena. In addition, the performance of capacity prediction gradually improves as the number of activated experts increases from $k=1$ to $k=3$. However, when $k=4$, the MAE of capacity prediction increases to 0.0088, and the RMSE increases to 0.0193. This indicates that excessively increasing the number of experts and enhancing the diversity of scales leads to an increase in redundant information, which reduces the accuracy of prediction. This result also demonstrates the effectiveness of the proposed selective activation of patch-based MLP blocks. By selectively activating the top-k experts, we can effectively avoid this problem. For B0018 with large and frequent capacity regenerations, the performance is different from the other three batteries. When the number of activated experts is only $k=1$, the MAE is 0.0136 and the RMSE is 0.026, which is the best. When the number of activated experts increases from $k=1$ to $k=4$, the MAE and RMSE both gradually increase. This is because the magnitude of capacity regeneration of B0018 is large. Increasing the diversity of time scales leads to an increase in the redundant features captured, affecting the prediction performance. In short, the adopted selective activation strategy enables flexible adjustments for battery capacity sequences with different characteristics, resulting in superior feature extraction and capacity prediction with excellent interpretability.

\subsection{Effects of intra-patch MLP and inter-patch MLP}
To validate the effectiveness of the proposed patch-based MLP block, we also evaluate the capacity prediction performance when either the intra-patch MLP or the inter-patch MLP is removed. As shown in \autoref{tab_2} and \autoref{fig_13}, the average MAE of capacity prediction without the intra-patch MLP increases to 0.0089, and consequently the average RMSE increases to 0.018, demonstrating the limitation of lacking local details. Similarly, the removal of the inter-patch MLP also results in an increase in the prediction error, suggesting the necessity of modeling global dependencies. This result confirms the effectiveness of the designed model, which employs intra-patch MLP to capture local details and inter-patch MLP to capture global dependencies.

\subsection{Inference time}
To assess the practicality of our method for battery capacity prediction, we compare the inference time of the four best-performing methods. As shown in \autoref{tab_3}, while our proposed method exhibits a slightly longer inference time, the cost is relatively small for performance improvement. Additionally, all methods demonstrate extremely fast inference speeds, well below 0.002 seconds, making them suitable for real-time applications.

\begin{table}
\centering
\begin{tabular}{l c c c r}
\hline
Method & ASTLSTM & AttMoE & EMD+LSTM+GPR & Proposed \\
\hline
Inference Time (s) & 0.0016 & 0.0009 & 0.0008 & 0.0017 \\
\hline
\end{tabular}
\caption{Average inference time for single capacity prediction using different methods}\label{tab_3}
\end{table}

\section{Conclusion} \label{Section 5}
In this paper, a multi-scale model, MSPMLP, is proposed to address the challenges posed by capacity regeneration phenomena and complex operating conditions for lithium-ion battery capacity prediction. In MSPMLP, patch-based MLP blocks are utilized to capture the long-term trends and local features in capacity sequences. Besides, MoE is employed to fuse features captured under different patch sizes, thereby obtaining multi-scale features to enhance the expressiveness and adaptability. Experimental results demonstrate that MSPMLP exhibits high accuracy and robust generalization, with a 41.8\% improvement in MAE for capacity prediction compared to existing methods. Visualization of prediction results and tests on the impact of each module of MSPMLP further confirm the efficacy of the proposed model. Future investigations will prioritize the evaluation and deployment of the proposed model on real-world operational data, addressing the deficiencies in data volume and diversity of operating conditions of current experimental dataset.

\section*{CRediT authorship contribution statement}
Yuzhu Lei: Methodology, Investigation, Software, Validation, Writing – original draft.
Guanding Yu: Methodology, Conceptualization, Validation, Writing- reviewing.

\section*{Declaration of competing interest}
The authors declare that they have no known competing financial interests or personal relationships that could have appeared to influence the work reported in this paper.

\section*{Data availability}
Data will be made available on request.

\section*{Acknowledgments}
This work was supported by the “Pioneer” and “Leading Goose” R\&D Program of Zhejiang under grant No. 2024C01183.
 
\bibliography{elsarticle-template}

\end{document}